\documentclass[12pt]{article}

\usepackage{multirow}
\usepackage[table,xcdraw,booktabs]{xcolor}
\usepackage{graphicx}
\usepackage{caption}
\usepackage{times}
\usepackage{booktabs}
\usepackage{amsfonts}
\usepackage{titlesec}
\usepackage{ulem}
\usepackage{natbib}

\graphicspath{{./figH/}}

\begin{document} 

\title{A foundation model enpowered by a multi-modal prompt engine for universal seismic geobody interpretation across surveys}


\author
{Hang Gao$^{1}$, Xinming Wu$^{1\ast}$, Luming Liang$^{2\ast}$, Hanlin Sheng$^{1}$,\\
 Xu Si$^{1}$, Hui Gao$^{1}$ and Yaxing Li$^{1}$\\
\normalsize{$^{1}$School of Earth and Space Sciences, University of Science and Technology of China,}\\
\normalsize{Hefei, China.}\\
\normalsize{$^{2}$Applied Sciences Group, Microsoft}\\
\normalsize{Bellevue, US.}\\
\\
\normalsize{$^\ast$To whom correspondence should be addressed:}\\
\normalsize{E-mail: xinmwu@ustc.edu.cn}
}
\date{}






\maketitle

\begin{abstract}
 Seismic geobody interpretation is crucial for structural geology studies
and various engineering applications. Existing deep learning methods
show promise but lack support for multi-modal inputs and struggle
to generalize to different geobody types or surveys. We introduce a
promptable foundation model for interpreting any geobodies across
seismic surveys. This model integrates a pre-trained vision foundation model (VFM) with a sophisticated multi-modal prompt engine.
The VFM, pre-trained on massive natural images and fine-tuned on
seismic data, provides robust feature extraction for cross-survey generalization. 
The prompt engine incorporates multi-modal prior information to iteratively refine geobody delineation. 
Extensive experiments
demonstrate the model’s superior accuracy, scalability from 2D to 3D, 
and generalizability to various geobody types, including those
unseen during training. 
To our knowledge, this is the first highly
scalable and versatile multi-modal foundation model capable of interpreting any geobodies across surveys 
while supporting real-time interactions. 
Our approach establishes a new paradigm for geoscientific
data interpretation, with broad potential for transfer to other tasks.
\end{abstract}

\section*{Introduction}\label{sec1}

Geobody interpretation is a fundamental operation in the field of seismic exploration,
entailing the identification and segmentation of
geobodies across seismic data \citep{Klausen2021}.
The achievement of precise geobody segmentation is crucial for various geological
applications, such as well system design, 
petroleum development and geological modelling \citep{Olaniyi2019}.
For instance, delineating channel within seismic data aids
in pinpointing prospective hydrocarbon reservoir,
guiding the discovery and exploitation of oil and
gas fields \citep[]{bridge2000interpreting,payenberg2003reservoir}.
Interpreting salt bodies contributes to a 
deeper understanding of salt tectonics, 
crucial for delineating the hydrocarbon traps \citep{jackson2017salt}.
The detection of paleokarsts facilitates the study of 
paleokarst system, forming carbonate reservoirs \citep{dou2011paleokarst}.
Traditionally, manual segmentation has been considered the benchmark for accurately delineating geological bodies within seismic data. 
However, this approach is time-consuming, labor-intensive, and requires a high level of expertise \citep[]{wu2019faultseg3d,shi2020}.
The subjectivity of manual interpretation also poses challenges 
to the consistency of geobody segmentation.
To reduce the reliance on manual labor, 
various seismic attributes, such as coherence
\citep[]{marfurt1998suppression,wu2017directional,liao2019analysis},
variance \citep[]{marfurt1998suppression,abul2012subsurface}
, curvature \citep[]{chopra2007volumetric,chopra2007seismic,alaei2017seismic}
and fault enhancement \citep[]{cunningham2019}
are proposed to assist researchers in
delineating geobodies in seismic data \citep{ALVARENGA202261}.
Nonetheless, these attributes are often susceptible to field noise and complex strata \citep[]{marfurt2014,shi2019salt},
making it challenging to refine geobodies.

Recently, the application of convolutional neural network (CNN)
from deep learning domain 
has seen great success in multiple computer vision tasks, 
such as image segmentation \citep[]{ronneberger2015u,badrinarayanan2017segnet,chen2017deeplab},
object detection \citep[]{girshick2015fast,ren2015faster,redmon2016you}
and instance segmentation \citep[]{dai2016instance,he2017mask,bolya2019yolact}.
This success has prompted many researchers to 
consider geobody interpretation as 
a semantic segmentation task and 
employ a CNN to delineate specific geobody \citep{mousavi2020earthquake},
such as
salt body \citep[]{waldeland2018,di2018deep,shi2019salt,di2020comparison},
horizon \citep[]{tschannen2020extract,luo2023seq},
paleokarst \citep{wu2020deep},
seismic facies \citep{zhao2018seismic},
fault \citep[]{zhao2018fault,wu2019faultseg3d,gao2021fault,li2022automatic,bonke2024data}
and oceanic gas hydrate
\citep{geng2020automated}.
CNN-based methods have emerged as transformative approaches,
significantly alleviating the time and labor burdens of traditional methods \citep{wu2020building}.
However, a significant limitation that plagues these CNN-based models is their inherent task-specific nature.
These models are typically trained for particular geobody segmentation and therefore lack the capability of interpreting different types of geobodies, increasing the complexity of the interpretation process
and the difficulty of their deployment.
In addition, the inductive bias of CNNs make them experience a substantial performance decline in field seismic data \citep{dosovitskiy2020AnII}.
This poses a considerable generalization barrier, hindering
their broader integration into universal seismic interpretation.
Handling data from different surveys often necessitates repeated dataset construction and model training,
significantly limiting their application.
Moreover, existing CNN-based models face significant challenges in incorporating geological priors, expert knowledge, and other multi-modal information, 
making it difficult to iteratively refine their predictions. 

The advancement of in Vision Foundation Models (VFMs) holds promise for resolving these issues. 
This approach aims to train a VFM  
with massive parameters 
using a large-scale image datasets 
to achieve feature extraction in universal images.
Notable examples of VFMs include 
CLIP \citep{radford2021learning},
MAE \citep{he2022masked}
and
DINOv2 \citep{oquab2023dinov2}. 
These models exhibit exceptional adaptability and excellence in various natural images, 
highlighting a paradigm shift 
towards more universally applicable foundation models in visual tasks.
This paradigm motivate some researchers to
construct vision foundation models for geoscience \citep[]{li2023tariq,sheng2023sfm}.
These foundation models trained on large-scale datasets
are applicable to various downstream geoscientific tasks, such as seismic interpretation, denoising, and interpolation.
While these models have achieved excellent performance in seismic data processing tasks,
they still require independent fine-tuning for each downstream task and especially fail to support multi-modal interactions that are often necessary for seismic geobody interpretation.

The development of multi-modal foundation models 
like SAM \citep{Kirillov2023SegmentA} 
brings novel solutions to this problem. 
SAM can integrate various types of prompts to segment targets, exemplifying a prompt-based interactive segmentation method that 
points or bounding boxes to delineate objects of interest. 
Compared to traditional CNN-based methods, 
SAM exhibits greater generalizability and applicability. 
Recently, researchers have ventured into the deployment of SAM models to medical image segmentation~\citep{ma2024segment}, remote sensing image segmentation \citep[]{yan2023mosam,zhou2024mesam} and crater detection \citep{giannakis2024115797}.
These methods inspire us to develop a universal geobody interpretation model. 
Such models can be pre-trained on a large-scale datasets 
and be applied to seismic interpretation tasks 
across multiple types of geobodies.  
However, due to the differences 
between seismic data and natural images,
directly applying SAM without training
to field seismic data has yielded suboptimal results. 
Fundamentally, the trainset of pre-trained SAM
lacks image-mask pairs related to seismic data, 
rendering the model ineffective in characterizing seismic images. 
Unlike the success of MedSAM in medical images \citep{ma2024segment}, 
seismic interpretation domain currently 
lacks a massive annotated seismic dataset for training and validation. 
Therefore, retraining a SAM specifically for seismic data 
is extremely costly in terms of computation and data annotation.

\begin{figure}[t!]
	\centering	\includegraphics[width=\columnwidth]{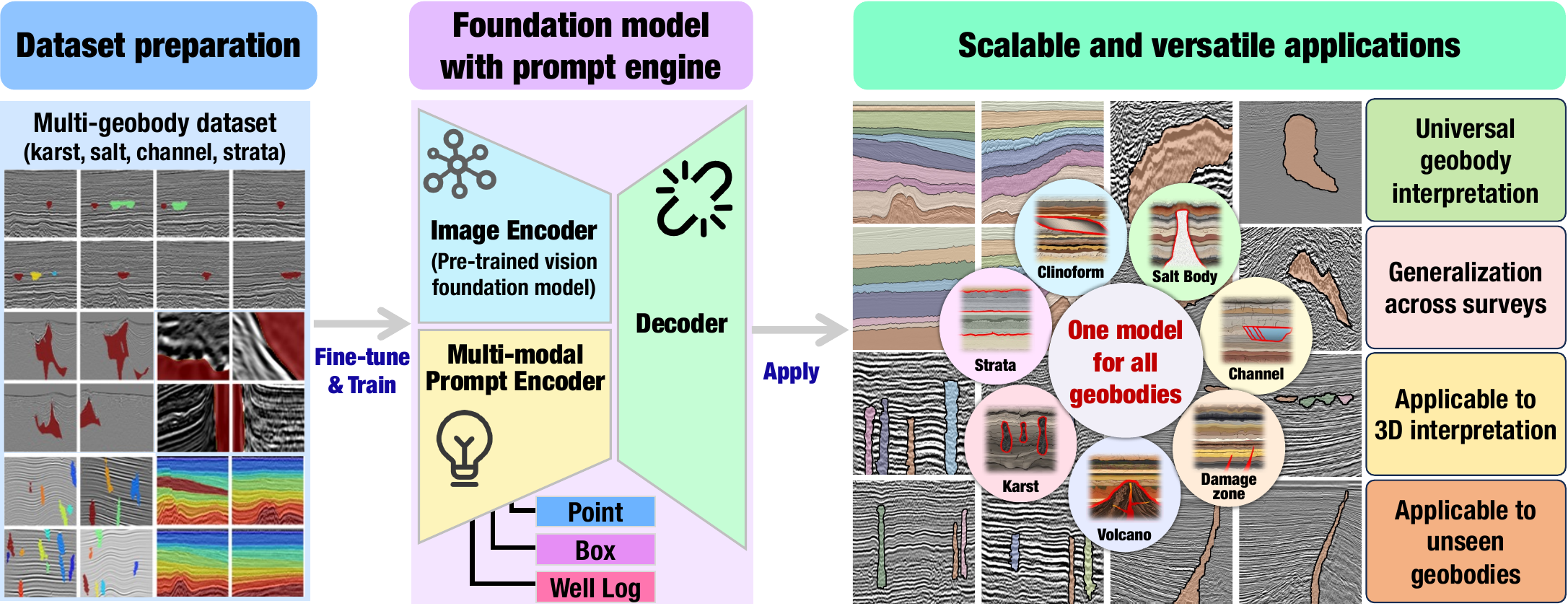}
	\caption{
		Workflow of developing our model of segment any geobodies (SAG) across surveys. 
		We start by collecting diverse seismic images and labeling various geobodies to construct a multi-geobody dataset (first column). 
		We further integrate a pre-trained foundation model
		with a well-designed prompting engine and decoder (second column), which are fine-tuned and trained on the multi-geobody dataset.
		This approach results in a model that can achieve
		real-time, interactive segmentation of any geobodies across various seismic surveys (third column).
		Moreover, this model, without retraining, can be directly extended to interpret 3D geobodies and other geobody types unseen in the training dataset.
	}
	\label{fig:01}
	
\end{figure}

In this paper, we propose a workflow (Fig.~\ref{fig:01}) to 
develop a model of interpreting or segmenting any geobodies (SAG) 
with strong generalization capabilities across surveys. 
This is achieved by combining a pre-trained VFM with a multi-modal prompt engine, which serve as encoders to extract key features from input seismic data and multi-modal prompt information, respectively.
The VFM, pretrained on a 
massive-scale dataset of natural images, 
possesses the ability to extract and represent key features from various input data, 
providing a basis for cross-survey generalization. 
Its generalizability is adapted and transferred to seismic data through the parament-efficient fine-tuning with Low-Rank Adaptation (LoRA) \citep{hu2022lora}.
The multi-modal prompting engine incorporates prior information related to target geobodies, 
such as points, boxes, and well logs, derived from expert interpretations 
or other automated methods. 
This prompt-guided prediction process results in more geologically and 
geophysically consistent outputs, further enhancing the model's reliability and generalizability.

Our workflow (Fig.~\ref{fig:01}) of developing the promptable foundation model for geobody interpretation involves several key steps. 
First, we construct a dataset containing multiple geobody types (Fig.~\ref{fig:01}, first column). 
We then integrate a pre-trained foundation model with carefully designed prompting engine and decoder (Fig.~\ref{fig:01}, second column), followed by fine-tuning and training. 
This process results in a model capable of real-time, interactive 
segmentation of any geobodies across different seismic surveys (Fig.~\ref{fig:01}, third column). 
Notably, despite being trained on 2D data, our model can interpret 3D geobodies through sequential 2D predictions, recursively using previous predictions as prompts to maintain consistency. 
Furthermore, the model can detect geobodies like fault damage zones and volcano that do not present in the training dataset, enhancing its versatility and generalization. 
Our approach introduces a novel paradigm for develop applications of foundation models to geoscientific data analysis. 
\section*{Results}\label{sec2}
\begin{figure}[t!]
	\centering
	\includegraphics[width=0.65\columnwidth]{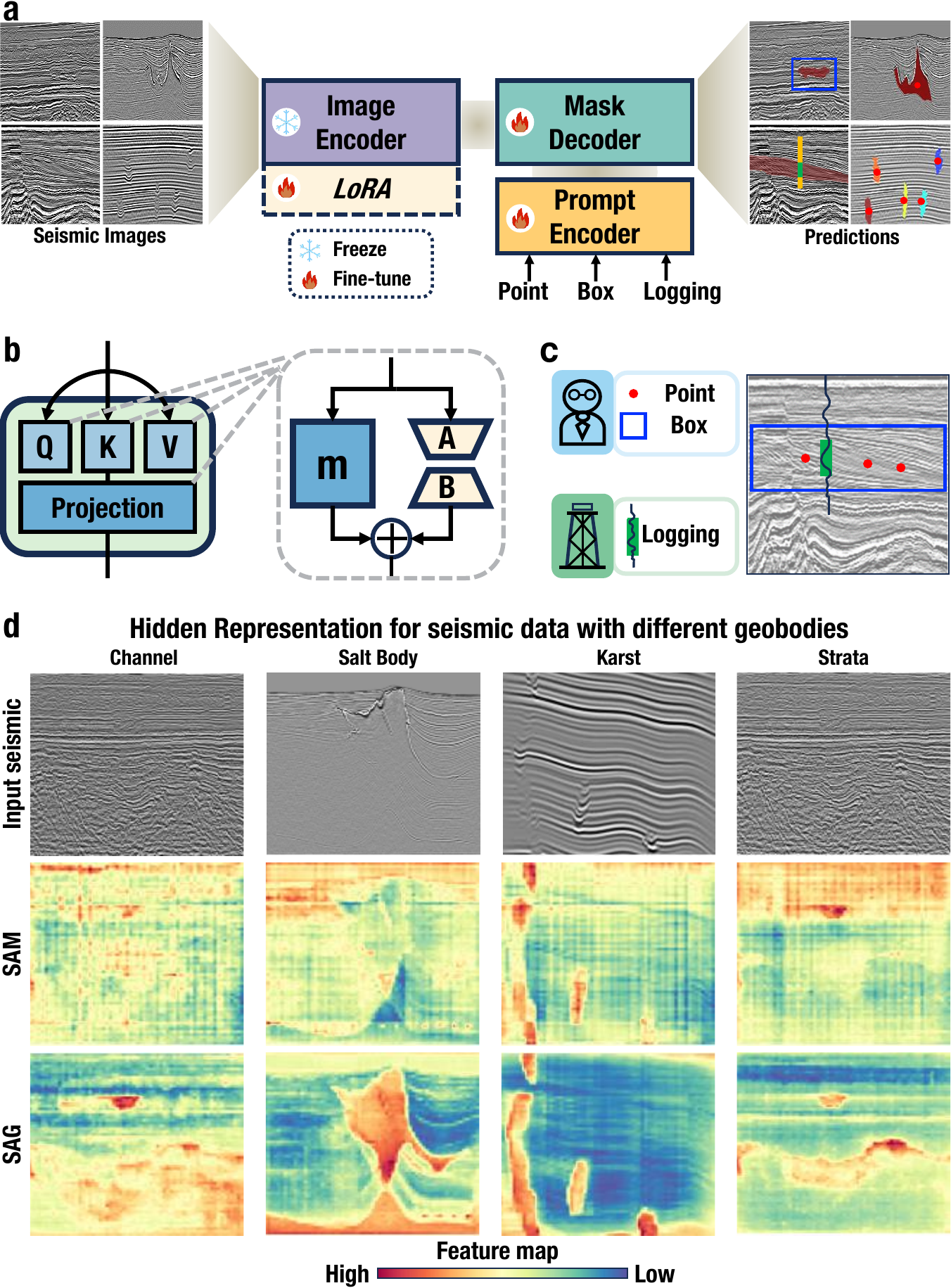}
	\caption{
		Network architecture, training strategy
		and interpretability of image encoder. 
		\textbf{a} 
		The SAG model includes an image encoder for 
		characterizing seismic data, 
		a prompt encoder for encoding diverse prompts, 
		and a mask decoder for generating masks. 
		During training, the pre-trained image encoder remains frozen, while its associated LoRA modules, prompt encoder, and mask decoder are trained.
		\textbf{b} LoRA module for fine-tuning the pre-trained model. LoRA adapters are inserted into each attention module within the image encoder, allowing fine-tuning of its parameters.
		\textbf{c} Point, box and well logging prompts inputted into prompt encoder
		\textbf{d} Visual interpretability analysis to 
		the hidden representations of seismic features 
		by the pre-trained model (SAM) and the fine-tuned model (SAG).
		Seismic images containing various geobodies (first row) 
		are input into SAM and SAG to 
		visualize their respective feature maps (second and third rows).
	}
	
	\label{fig:02}
\end{figure}

\subsection*{Data preparation}\label{sec2sub2}

To fine-tune the encoder of pre-trained VFM and to train the prompt engine and decoder in developing our SAG model, we build a multi-geobody dataset with publicly available field and synthetic seismic data. 
The synthetic data with karst labels were derived from the dataset published by~\citep{wu2020deep}.
The field data are collected from diverse public sources such as
United States Geological Survey (USGS), 
New Zealand Petroleum and Minerals (NZPM), Netherlands Oil and Gas (NLOG), 
Society of Exploration Geophysicists(SEG).
Many legacy seismic data,
acquired earlier,
suffer from poor imaging quality or empty areas.
We exclude these data 
to maintain the quality of dataset.
Additionally, most seismic data 
do not contain typical geobodies and the 
corresponding manual annotation.

We look though the collected seismic data and find these containing typical geobodies and then manually or semi-automatically annotate the geobodies 
to construct multi-geobody datasets.
These datasets, derived from various seismic surveys worldwide, 
encompass a wide range of subsurface geological features and a variety of typical geobodies (Supplementary Fig.~1). 
As depicted in Supplementary Fig.~2, 
we constructed a dataset with a total number of 25,000 samples 
containing three geobody types of paleokarsts (15,360 samples), 
salt bodies (3,507 samples), strata (3,464 samples)
and channels (3,234 samples) with various seismic expressions.
Unlike existing CNN-based category-level 
segmentation methods, 
SAG requires instance-level masks for fine-tuning.
Therefore, we transform 
these category-level masks into
instance-level masks,
ensuring each individual geobody has its
corresponding mask (Supplementary Fig.~3).
Based these instance-level masks,
we can construct various types of prompts 
(point, bounding box and well logging) for 
guiding the VFMs in segmenting 
the target geobody (Supplementary Fig.~4).
This multi-geobody dataset 
provide diverse seismic images,
geobody masks and the corresponding prompts
for the adaptive fine-tuning of pre-trained VFM and the training of 
the prompt encoder and decoder.

\subsection*{Overall architecture and interpretability analysis of SAG}\label{sec2sub3}
Our SAG model features an overall Encoder-Decoder network architecture (Fig.~\ref{fig:02}a and
Supplementary Fig.~5), 
which primarily consists of an encoder to integrate input information and 
a decoder to produce output masks. 
The encoder part includes 
an image encoder for extracting and representing  seismic data features in the hidden space and a prompt encoder for encoding various prompts into embedding features that are concatenated with the seismic features in the hidden space. 
The image encoder is inherited from the large ViT encoder of the SAM pre-trained on massive natural images. 
To preserve the feature extraction capability from pre-trained encoder, we retain its pre-trained parameters during the continued training with our seismic dataset.
Simultaneously, to enable the model to accommodate seismic data,
we add a parallel LoRA module to each transformer layer of the image encoder 
for incremental fine-tuning (Fig.~\ref{fig:02}b). 
The prompt encoder is specifically redesigned to effectively adapt to specific prompts such as points, bounding boxes, and well logs (Fig.~\ref{fig:02}c) in the context of geobody interpretation.
The mask decoder, designed with cross-attention transformer, fusions the encoded seismic and prompt features and projects them to a probability map highlighting target geobodies.
Detailed discussion about the network architecture is included in the Method section and more detail of the pre-trained encoder, LoRA module, prompt encoder, and decoder is illustrated in Supplementary Fig.~5.
During the continued training with our multi-geobody dataset, the image encoder is fine-tuned with LoRA while the prompt encoder and decoder 
are fully trained, resulting in our SAG model. 
This model enables real-time, interactive segmentation of target geobodies with diverse geological prompts.

\begin{figure}[t!]
	\centering
	\includegraphics[width=0.8\columnwidth]{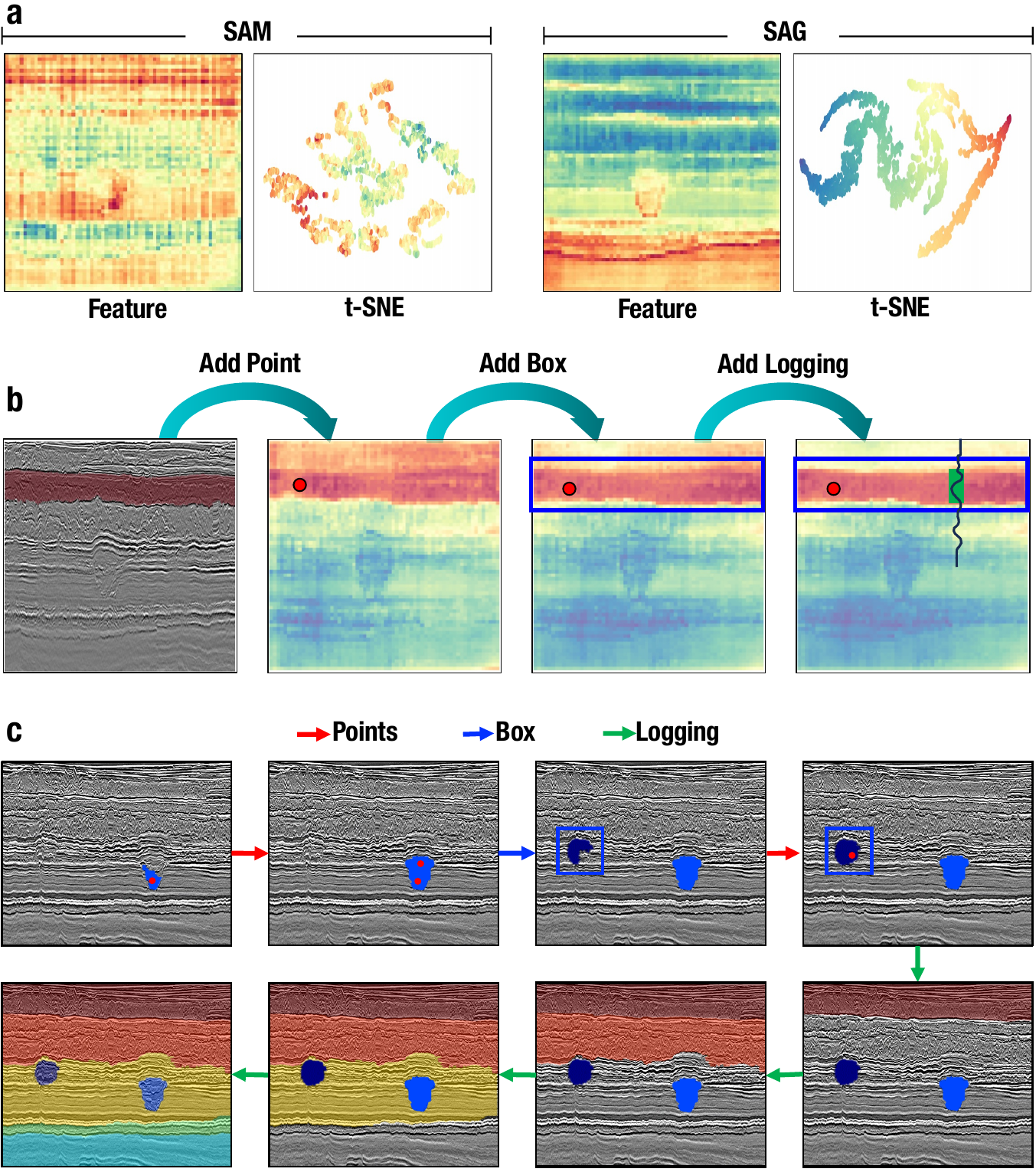}
	\caption{
		Visual interpretability analysis to the performance of SAG and its prompt engine. 
		\textbf{a} t-SNE dimensionality reduction of hidden representation from SAM and SAG. t-SNE analysis is a dimensionality reduction technique primarily used for visualizing high-dimensional data.
		\textbf{b} Impact of different prompts on features space of mask decoder.
		Iterative updating of prompts helps 
		the decoder to 
		obtain more accurate hidden representations of target geobody. 
		\textbf{c} Workflow of interpreting multi-type and multi-instance geobodies in a field seismic image with a single model of SAG.
	}
	\label{fig:03}
\end{figure}

As a state-of-the-art segmentation foundation model,
SAM performs well on natural image datasets 
but is challenging to apply directly to seismic geobody segmentation. 
Our initial experiments show that SAM can identify the main parts of target geobody in seismic images, demonstrating considerable generalization ability, however, it struggles to further refine the segmentation of geobodies.
To visually interpret and understand the limitation of direct application of SAM to seismic data, we input seismic images containing different types of geobodies (the first row of Fig.~\ref{fig:02}d) into the SAM's image encoder module and display the extracted feature maps in the second row of Fig.~\ref{fig:02}d, where some main features of the geobodies are apparent but their details and boundaries are not clearly characterized. 
In contrast, the image encoder in our fine-tuned SAG model can capture precise features 
related to geobodies from seismic images 
(the third row of Fig.~\ref{fig:02}d).
We further conducted a t-SNE (t-Distributed Stochastic Neighbor Embedding) analysis \citep{van2008visualizing} of the feature maps 
on a complex seismic image containing channel and multiple strata, as illustrated in Fig.~\ref{fig:03}a. 
The results indicate that, 
compared to the SAG model, SAM's hidden representations of seismic images 
are not spatially continuous, 
limiting its performance in geobody interpretation within seismic data.
SAG effectively represents different geobodies as specific clustered characteristics in the feature space, 
which facilitates the mask decoder to transform these features into high-quality masks highlighting geobodies.

To further explore the impact of different prompts on 
the interpretation capabilities of SAG, 
we visualize the feature representations of 
the mask decoder module under different prompt constraints in Fig.~\ref{fig:03}b. 
When we add only a single prompt point on seismic image
(the first panel of Fig.~\ref{fig:03}b), 
the top of seismic image 
exhibited high-value regional features (red in the second panel of Fig.~\ref{fig:03}b), 
indicating that a single prompt point is insufficient for accurate strata segmentation. 
We further use a bounding box to 
constrain the target strata, 
the high-value range is restricted to a specific area (red in the third panel of Fig.~\ref{fig:03}b), but the stratigraphic boundaries still remain unclear in some areas.
Finally, by adding lithological segments from well log, we obtain the high-value feature representation with clear and geologically reasonable stratigraphic boundaries (red in the forth panel of Fig.~\ref{fig:03}b).
This indicates that diversified mixed prompts effectively improve the characterization of geobodies from different perspectives, aiding in achieving interactive and refined geobody interpretations.

In summary, by displaying the feature maps of the SAG's encoder and decoder as shown in Fig.~\ref{fig:02}d and~\ref{fig:03}b, respectively, we provide a visual interpretation and understanding of the performance of our SAG. 
Compared to SAM without fine-tuning, 
SAG can effectively capture characteristics 
of various geobodies and further 
adjust the features based on prompts. 
These effective features form the basis 
for SAG to achieve universal segmentation 
of any geobodies across surveys, and improves the geological consistence of the SAG's predictions.

\subsection*{Prompt engine for multi-modal interactions}\label{sec2sub4}

The prompt encoder plays a crucial role in the network architecture, 
responsible for converting prompts into constraint information that 
guides the model in segmentation tasks. 
Consequently, it is essential to design the prompt 
encoder and the corresponding input prompts
specifically for geobody interpretation tasks. 
The originally pre-trained SAM offers three types of prompts: 
points, bounding boxes, and dense prompts 
(text prompts are currently not available in original codes). 
Among these, the point and bounding box prompts 
support human-computer interaction and 
can be used by an interpreter to guide the model 
in segmenting the target geobody. 
However, dense prompts are often overlooked by researchers because they originate from additional intermediate features within the image encoder module, making them difficult for users to directly constrain the target. 
This limitation in prompt encoder hinders the SAM-based method from fully leveraging multi-modal information to segment geobodies.  

To address this issue, we design a new type of prompt specifically for geobody interpretation: well logging prompts. 
In the actual seismic interpretation process, 
we use well logs as a reference for geobody interpretation. 
Lithological interpretations from core samples or logging curves provide us with ground truth of the subsurface, which can be viewed as weakly-supervised labels, offering local rather than global supervision information. 
While such logging data can be used as segmentation labels 
through weakly-supervised learning methods, 
they are difficult to be employed as flexible and scalable prompts 
to iteratively constrain geobody segmentation. 
As shown in Supplementary Fig.~3, 
we convert the lithological segments from logs 
into binary masks and use them to constrain the geobody segmentation. 
In practical seismic interpretation, this well-log constraint is especially useful for interpreting channels and strata. 

During the training process of SAG, 
point, bounding box, and 
well logging prompts are required to 
guide the model in segmenting geobodies. 
As depicted in Supplementary Fig.~3, 
we construct three types of prompts 
based on instance-level labels 
from multi-geobody dataset. 
First, we randomly select a small number of points within the mask region to 
serve as point prompts. 
Then, we calculated the bounding box around the mask region in the binary label 
and expanded it randomly to simulate perturbations 
in actual seismic interpretation. 
Finally, we constructed logging prompts 
by randomly extracting multiple vertical strip labels from the masks. 
These prompts play a crucial role in constraining the SAG model during interpretation. 
Using the flexible combinations of these prompts as guidance, our SAG model is able to fully interpret all the strata and channel bodies within a same seismic image as shown in Fig.~\ref{fig:03}c. 
This general capability of SAG to adaptively interpret varying geological body types and instances within the same seismic image surpasses all previous traditional algorithms and deep learning methods.
The specialized design of the prompt encoder 
and multimodal prompts
makes SAG
more scalable
for universal geobody interpretation.


\subsection*{Scalability and versatility of SAG }\label{sec2sub5}

\begin{figure}[t!]
	\centering
	\includegraphics[width=0.8\columnwidth]{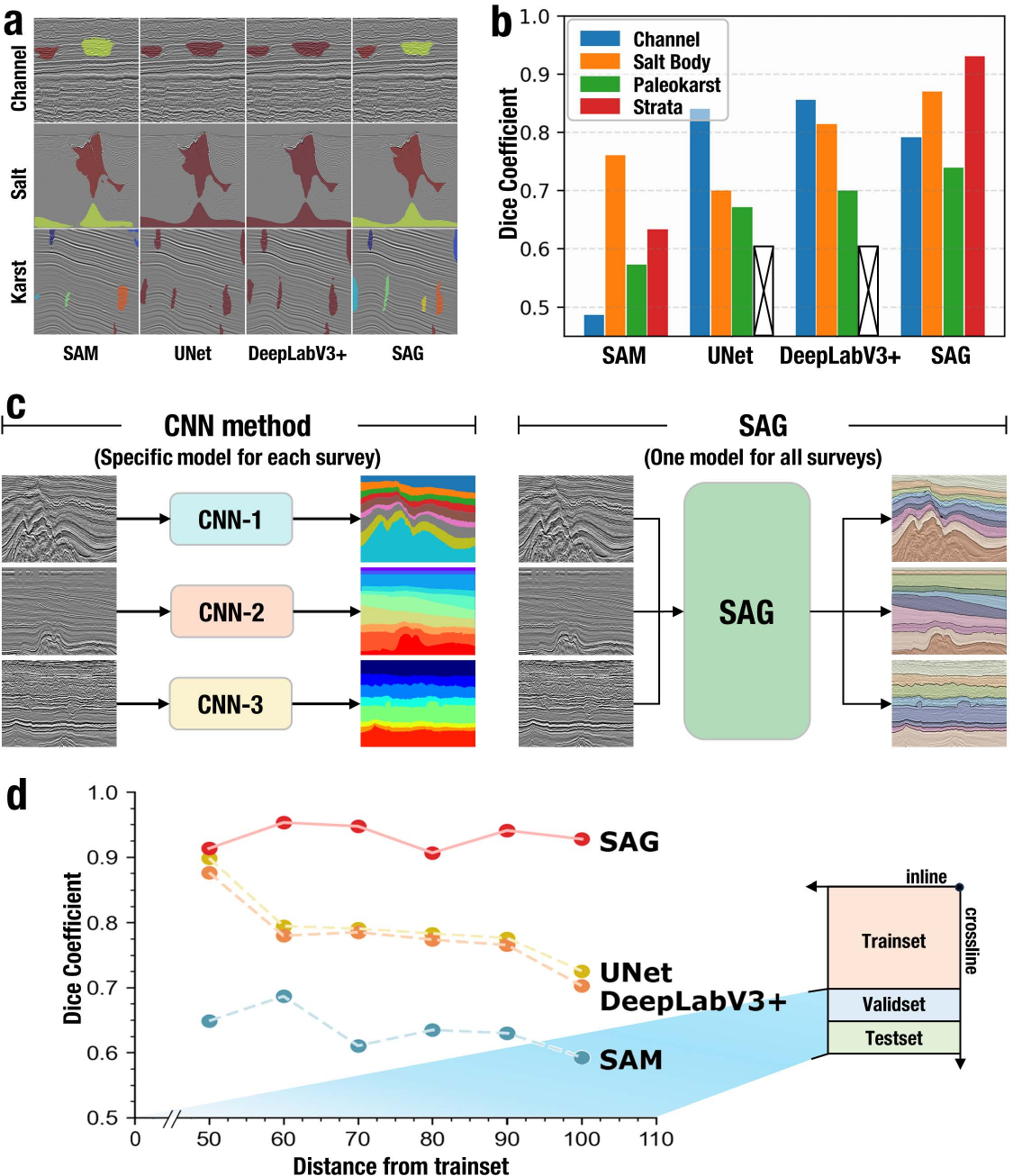}
	\caption{Quantitative and comparative evaluation on test datasets. 
		\textbf{a} Visualization of multi-geobody predictions by different models on the test set. 
		\textbf{b} Evaluation metrics on various geobody interpretation by different models. 
		\textbf{c} Limitations of CNN-based methods for strata interpretation across different seismic data. 
		The CNN-based approaches require training a specialized model for each seismic survey to interpret its strata, whereas a single model of SAG can achieve strata segmentation across various surveys.
		\textbf{d} Comparison of model performance variation with distance from the trainset.
		The difference between 
		test samples and trainset 
		increases with distance.
	}
	\label{fig:04}
\end{figure}

\begin{figure}[t!]
	\centering
	\includegraphics[width=0.8\columnwidth]{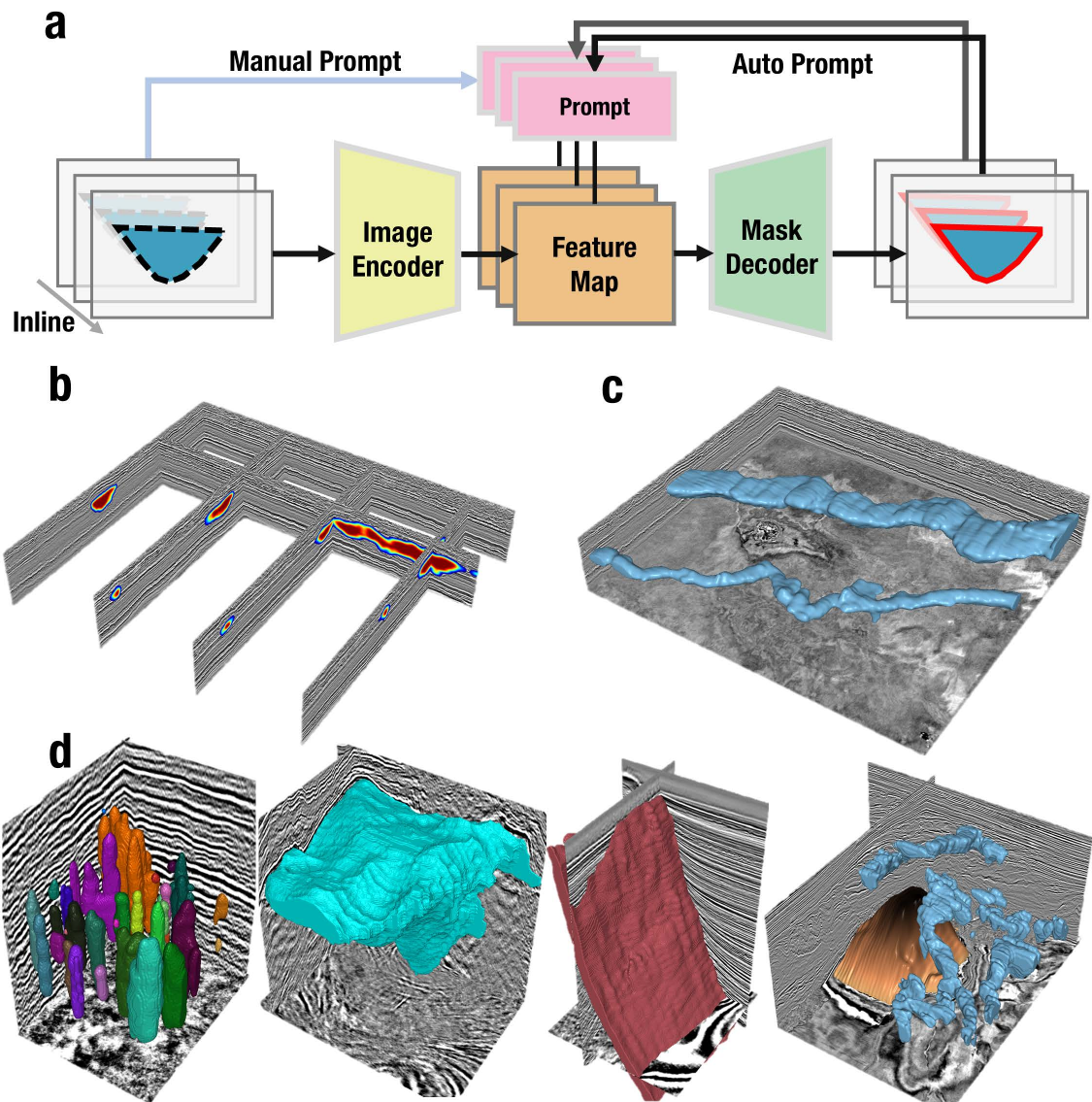}
	\caption{
		Application of SAG model to interpret 3D geobodies and other geobody types unseen in training dataset.
		\textbf{a} Implementation of 3D geobody interpretation by sequential 2D predictions and recursively use previous predictions to automatically generate prompts for guiding the next predictions to maintain consistency.
		\textbf{b} Visualization of 3D Channel likelihood volume predicted by SAG. 
		\textbf{c} Modeling of a 3D deep-water channel system from the likelihood volume.
		\textbf{d} 3D Interpretation of instant paleokarsts, salt body, fault damage zone, volcanic complex and overlying channel system.}
	\label{fig:05}
\end{figure}

Accurate geobody interpretation typically involves 
the integration and analysis of various geological and geophysical information, posing challengings to 
the scalability and versatility
of the existing deep learning methods.
A highly scalable geobody interpretation method should be able to generalize across seismic data from various surveys and allow the integration of various multi-modal constraint information. 
The image encoder and prompt encoder of SAG ensure this level of scalability.
The image encoder of SAG, inherited 
from a VFM pre-trained on massive natural images, possesses the robust ability of rich feature representation to various input images.
This indicates that SAG is directly scalable to interpret geobodies in natural image formats, such as outcrop images and physical modeling images.
After fine-tuning on our seismic geobody dataset, its generalizability across various natural images is adapted and transferred to various seismic data across surveys.
Extensive experiments across more than ten different surveys (Fig.~\ref{fig:01},~\ref{fig:04},,~\ref{fig:05} and Supplementary Fig.~6) demonstrate that SAG is scalable to interpret geobodies in seismic data with varying sampling rates, frequency spectra, signal-to-noise ratios, geologic patterns and so on.
Moreover, its stable performance across diverse seismic data enables training on 2D data while performing recursive inferences for 3D geobody interpretation, thereby allowing its scalability from 2D to 3D.
Finally, The combination of the image encoder's universal feature extraction 
with expert guidance based on the prompt engine 
enables SAG to be scalable for interpreting other types of geobodies not seen in the training data.

The versatility design of SAG is reflected in its capability for 
instance-level universal segmentation of multiple geobodies. 
The incremental fine-tuning of
the pre-trained VFM enables SAG
to adapt to 
different geobodies and seismic surveys.
And the decoupled design of the image encoder
and prompt encoder in SAG
allows us to interpret different geobodies
with various prompts
in the same seismic image.
The image features from image encoder 
and prompt embedding from prompt encoder 
are separately input into the mask decoder. 
This allows for seismic image features to be computed only once and 
reused for interpreting multiple geobodies within the same image. 
By combining the same image features with different prompts, 
multiple geobody interpretations can be achieved continuously. 
This greatly reduces the number of feature extraction processes 
required for interpreting multiple geobodies, 
effectively lowering computational costs and increasing the model's inference speed.
This feature allow us
to achieve versatile segmentation of
geobodies on diverse seismic data
using only a single model of SAG.
As shown in Supplementary Fig.~6, 
SAG can quickly and easily segment
diverse types of geobodies
across different field seismic surveys.


Scalability and versatility of SAG 
can improve the efficiency of geobody interpretation.
To demonstrate this, 
we conducted a geobody interpretation in the Romney3D field case from New Zealand. 
As shown in Fig.~\ref{fig:03}c and video in supplementary material, we interpret a seismic image 
containing two deep-water channels 
and multiple strata. 
Initially, using a single prompt point, we only obtain the basal part of the right channel body. 
To interpret the entire deep-water channel, 
we add two more prompt points above the basal part, successfully delineating the entire channel body. 
For the left channel, 
we use a bounding box as a prompt to interpret part of the channel body and then refine the boundaries 
with a control point to achieve a complete interpretation. 
Finally, we use well log prompts to 
interpret multiple strata.
In this way, a single SAG model achieves the full interpretation of multiple geobodies in a seismic image, eliminating the need to train multiple specialized models, one for each geobody type, as required by traditional deep learning methods.
In addition, we iteratively use
various prompts to 
refine the boundaries of geobodies, thereby
achieving more accurate interpretation.
Traditional deep learning-based methods 
require retraining to update or correct its predictions.
By adopting the prompt-based
intelligent interpretation paradigm,
SAG has achieved controllable segmentation of multiple geobodies
and greatly improved the efficiency
of seismic interpretation.

\subsection*{Quantitative and qualitative analysis}\label{sec2sub6}

We evaluated the SAG model through four geobody interpretation tasks: channel, salt body, paleokarst, and strata. 
Specifically, we compared it 
with SAM and CNN models
that frequently applied in seismic interpretation,
such as UNet \citep{ronneberger2015u} and DeepLabV3+ \citep{chen2017deeplab}. 
The UNet and DeepLabV3+ models 
were separately trained and evaluated 
on the specific geobodies 
of channels, salt bodies, and paleokarst. 
In contrast, SAM and SAG models performed instance-level universal segmentation on various geobodies under prompt constraints, 
and their performances were categorized and evaluated accordingly.
Fig.~\ref{fig:04}a visualizes the multi-geobody
segmentation of four models.
It can be found that SAG performed more
accurate instance-level segmentation.
The UNet and DeepLabV3+ models achieved
suboptimal performance in category-level 
geobody segmentation.
The SAM model, due to the lack of fine-tuning, 
exhibited inferior segmentation performance.
The statistical results in
Fig.~\ref{fig:04}b and Supplementary table.~1
indicate that, except for the channel
interpretation task,
SAG surpasses the UNet and DeepLabV3+ models
in multi-geobody interpretation tasks.
The good performance of the UNet and DeepLabV3+ models in the channel task 
might be due to the fixed locations of channel 
in the train and test set.
These location information can be easily learned by the inductive bias of CNN networks,
but limit their generalization.
SAG, enpowered by a multi-modal prompt engine,
enable researchers to interpret multiple geobodies
at different locations in different seismic surveys.
The evaluations of strata segmentation task
for UNet and DeepLabV3+ models are missing.
Because the fixed output of network
requires CNN-based methods (UNet and DeepLabV3+)
to train various models for strata in different
seismic surveys (Fig.~\ref{fig:04}c).
This severely restricts the application 
of UNet and DeepLabV3+ models to
strata segmentation across different surveys.
Therefore, we only evaluated the performance of 
SAG and SAM models in the strata interpretation task, 
and the results show that 
the SAG model achieves significantly better performance 
in the strata segmentation task.

To further investigate the generalization ability 
of these models to new data, 
we selected the Romney3D seismic dataset for evaluation. 
We divided the Romney3D seismic surveys 
into consecutive training, 
validation, and test sets 
based on spatial location. 
We trained and validated UNet, DeepLabV3+, and SAG individually 
on the training and validation sets. 
Subsequently, the four models (UNet, DeepLabV3+, SAG, and the unfine-tuned SAM) 
were used to predict the test set separately. 
Finally, we plotted the test results as line statistics (Fig.~\ref{fig:04}d) 
according to the spatial distance between 
the test samples and the training set. 
As shown in Fig.~\ref{fig:04}d, 
SAG maintains stable performance as the spatial distance 
from the training set increases 
(indicating an increase in the difference 
between the test data and the training set). 
In contrast, UNet and DeepLabV3+ exhibit significant decreasing performance 
with increasing distance due to the intrinsic inductive bias in CNN models. 
This demonstrates that the SAG model design ensures stability on unseen data, 
enhancing its potential extrapolation prediction capability in 3D seismic data.

The quantitative analysis indicates
that SAG has stronger
representation ability of 
seismic image
and generalization ability across
geobody category and seismic data.
This makes it possible for us to 
apply SAG to
interpret multiple geobodies of
arbitrary seismic data.
Researchers no longer need to 
train multiple specialized neural
networks to predict various 
geobodies from different seismic surveys 
and then aggregate them
for comprehensive geological structure analysis,
but instead simply use a single of SAG to 
complete all of these tasks.

\subsection*{The potential of extending SAG to 3D seismic data and other geobodies}\label{sec2sub7}
The high stability of SAG's extrapolated predictions (Fig.~\ref{fig:04}d) demonstrates its capability to extend to 3D geobody interpretation.
Specifically, this stable extrapolative predictive capability allows us to transform the 3D geobody delineation into
a sequence of 2D iterative predictions\citep{liu2024segmentmedicalmodelextended}.
In addition, the prompt engine of SAG allows us to recursively use the previous prediction to generate prompts for guiding the next prediction, which further helps maintain the consistency of extrapolative predictions.
As shown in Fig.~\ref{fig:05}a, to achieve 3D seismic interpretation, we decompose the 3D volume along the inline or crossline directions into a series of continuous 2D seismic profiles. 
We select one profile as initial
section and interpret its geobody with manual prompts.
Then we use the predicted mask of the initial section to automatically generate prompts for adjacent sections in the forward and backward directions.
By recursively predicting and prompting along a 
specific direction, 
we can achieve laterally consistent delineation of 3D geobodies in a 3D seismic volume.

Fig.~\ref{fig:05}b and~\ref{fig:05}c demonstrate the application of 
this method for 3D geological modeling of two deep-water channels in the Romney field case in New Zealand. 
We begin by manually prompting SAG to segment two channels in a middle section, initiating the recursive prediction process in both forward and backward directions. 
Then, we use automatic prompts from the previous section's prediction for guiding SAG to consistently segment the channel bodies in the subsequent sections.
After predicting all the 2D sections, 
we combine these results into a 3D channel likelihood volume (Fig.~\ref{fig:05}b) and build the 3D geometric modeling
of deep-water channel system (Fig.~\ref{fig:05}c) in this region.
Similarly, the SAG model has also been successfully applied to 
interpret 3D paleokarsts and salt bodies
in another 3D field seismic volumes 
as shown in the first and second panels of Fig.~\ref{fig:05}d, respectively.
More importantly, the SAG provides instance-level interpretation of 
geobodies (such as the individual paleokarsts in Fig.~\ref{fig:05}d) rather than 
the class-level segmentation achieved by previous CNN-based methods.
This instance-level results facilitate further quantitative analysis of the 3D geometry of each geobody.


To further validate SAG's interpretation ability
for unseen geobodies,
we conduct 3D predicting of 
fault damage zones and 
volcanic complex in another two seismic surveys.
Despite these geobodies 
are not included in the training data, 
our SAG model can still accurately delineate them as shown in the third and forth panels in Fig.~\ref{fig:05}d.
This further demonstrates SAG's exceptional scalability and generalizability for universal geobody interpretation. 
Such capabilities have not been exhibited by any existing automatic geobody interpretation methods.

\section*{Discussion}\label{sec3}
We propose SAG, a foundation model with a prompt engine, 
designed for universal seismic geobody interpretation across seismic surveys. 
SAG is fine-tuned on a curated multi-geobody dataset,
which includes various geobody labels such as channels, salt bodies, paleokarst, and strata.
SAG achieves an optimal balance between manual and intelligent interpretation, 
making it a interactive tool for universal seismic interpretation.

Through evaluations in 2D and 3D seismic data, 
the SAG model has demonstrated robust capabilities in interpreting diverse geobodies (even those unseen in training data) and strong generalization to 
new seismic data. 
Its performance surpasses that of existing models trained for particular geobody. 
By interpreting geobodies in seismic data,
SAG helps to achieve geobody modeling and structural geometric analysis, thereby facilitating geological research and petroleum exploration.
For example, SAG can effectively characterize 
deep-water channel system in 3D field seismic data within complex structural environments. 
Moreover, SAG introduces a new paradigm for the application of foundation models, 
adapting pre-trained VFMs for seismic interpretation by parameter-efficient fine-tuning. 
This paradigm can be extended to develop other geoscientific foundation models for other extensive applications.

Although SAG possesses strong geobody interpretation capabilities, it has certain limitations. 
Extensive seismic interpretation practices 
have shown that accurate characterization of many geobodies or structures 
needs to be conducted within 3D seismic data. 
3D seismic data can provide more precise structural features and additional perspectives, 
which enhance the accuracy of geobody interpretation. 
Developing a 3D universal geobody interpretation model 
requires a more diverse 3D annotated geobody dataset. 
While we have established a 2D multi-geobody dataset 
by collecting and annotating public data, 
this dataset serves as only a starting point. 
It requires supplementation with more diverse geobodies or geological patterns 
and development towards 3D datasets. 
We plan our 2D multi-geobody dataset publicly available to encourage further research and development.
Additionally, we aim to expand more types of prompts, 
such as text prompts, to enhance intelligent seismic interpretation.
Despite these limitations, SAG remains highly practical,
as it successfully extends the generalization and multimodality of pre-trained vision foundation model
to seismic interpretation, enabling interactive and universal geobody interpretation.

In conclusion, this research underscores the feasibility of adopting pre-trained VFMs to develop a universal interactive model for geobody interpretation, 
thus overcoming the limitations of 
existing geobody-specific methods. 
As an first interactive foundation model 
for seismic geobody interpretation, 
SAG effectively integrates 
expert knowledge, well log constraints 
to achieve a universal geobody interpretation, 
contributing significantly to seismic interpretation
and geological modeling.

\section*{Methods}\label{sec4}

\subsection*{Seismic data preprocessing}\label{sec4sub1}
The seismic data, originally formatted in SEGY, are converted binary format files to facilitate the training and inference. 
The amplitude of seismic data varied significantly 
due to differences in acquisition systems,
geological structures and imaging quality \citep{revelo2021acquisition}. 
In addition, some field seismic data may contain singular or null values. 
To handle these problems, 
we clipped the seismic amplitudes 
to the range between the 0.5th and 99.5th percentiles. 
In order to match the data format with the input of pre-trained SAM,
we apply min-max normalization to these seismic images and remapped the values to [0, 255].
Considering that most seismic interpretation studies 
use grayscale color for visualization, 
we employ the `gray' colormap from matplotlib as
the mapping function from amplitude to color scale. 
Since the pre-trained model 
requires three-channel input, 
we replicated the single-channel data across the channel dimension to obtain three-channel RGB seismic images.
This preprocessing workflow is applied to both training and inference seismic data, which helps improve the data consistency from different seismic surveys.

Based on the processed seismic images, 
we manually or semi-automatically 
annotated diverse geobodies within them to construct image-mask pairs for training the SAG.
To adapt to the input and output sizes
of the pre-trained image encoder in SAG,
seismic images and the corresponding masks
were cropped to a size of $1024\times 1024$.
After these steps, we construct a multi-geobody dataset (Supplementary Fig.~1 and~2) with an image size of $1024\times 1024\times 3$ and a mask size of $1024\times 1024\times 1$.
This standardized data preparation ensured quality and stability of training data,
providing a solid foundation for subsequent model training and validation.

\subsection*{Network architecture and fine-tuning method}\label{sec4sub2}


As shown in Fig.~\ref{fig:02}a, SAG is built on the architecture with two main modules of encoder and decoder.
The encoder module consists of an image encoder that 
extracts feature from seismic images and
a prompt encoder that transforms prompts into prompt embeddings.
The decoder integrates image features and prompt embeddings to generate masks for geobodies.
The image encoder is inherited from the pre-trained encoder of SAM-base. 
Although SAM provides large and huge versions of pre-training parameters \citep{Kirillov2023SegmentA}, 
considering the real-time requirements of 
interactive seismic interpretation, 
we adopt the base version SAM.
Specifically, the image encoder consists of 12 self-attention transformer blocks, 
each comprising a multi-head attention module and an MLP layer (Supplementary Fig.~5).

Compared to natural image segmentation, 
seismic interpretation has 
the unique advantage of incorporating 
lithological interpretations from well logging as constraints. 
To expand the forms of prompts, we modify the prompt encoder module by adding a new prompt from well logs.
The prompt encoder (Supplementary Fig.~5) includes a patch-embedding layer 
and a mask-downsampling module, 
which are used to process sparse prompts (points and boxes) and dense prompts 
(well logging), respectively. 
The patch-embedding module maps sparse prompts 
into sparse encoding containing prompt position information. 
Specifically, the points are represented as position embedding and two learnable embeddings that are used to indicate positive or negative points corresponding to the target or background. 
The bounding box is represented as an embedding pair to denote the position information of the upper-left and lower-right corners of the box. 
Then the points and box embeddings are merged into multiple 256-dimensional features. 
The Mask-downsampling module consists of three 2D convolutions, 
two LayerNorm layers, 
and two GELU activation layers. 
This module downsamples dense prompts and convert them 
into 256-dimensional embeddings.

The mask decoder contains 2 cross-attention transformer blocks and an upsampling layer
for generating predicted masks.
When seismic images and diverse prompts are input into SAG, the image ($1024\times1024\times3$) is 
divided into multiple $16\times16\times3$ patches 
upon entering the image encoder and then transformed into $64\times64$ hidden representations
by multiple transformer blocks. 
The prompt input is converted 
into multiple 256-dimensional vectorial embeddings 
by the prompt encoder.
The mask decoder fuses the hidden representations 
from the image encoder and 
the embedding vectors from the prompt encoder 
and outputs a $256\times256$ mask. 
After being passed through an activation function, 
the mask is upsampled to match the input image size using bilinear interpolation.

The key to adapting the pre-trained model to seismic data is 
to fine-tune its image encoder module which is used for
characterize the image features.
The image encoder of pre-trained SAM includes many dense layers with full-rank weight matrices,
which poses challenges to retrain the whole module. 
To fine-tune this module, we used the LoRA method to
update the dense layer of image encoder efficiently
\citep{hu2022lora}. 
For the dense layer parameters $W_0 \in \mathbb{R}^{n\times m}$, 
we add a extra learnable low-rank decomposition parameter $\Delta W$ 
to update the hidden representation.
The $\Delta W$ can be represented by two linear layers
$B$ and $A$, where $B\in \mathbb{R}^{n\times r}$ and
$A\in \mathbb{R}^{r\times m}$. 
The rank $r$ of $A$ and $B$ is much smaller than input dimension $m$
and output dimension $n$, and can be empirically taken as an exponent of 2.
The LoRA fine-tuning method can be expressed by the following:

\begin{equation}
	\left\{
	\begin{array}{ll}
		h = W_0 x+ \Delta W x, & \\
		\Delta W = BA,
	\end{array}	
	\right.
	\label{equ:lora}
\end{equation}

where $h$ is the hidden representation of the dense layer from image encoder.
During the fine-tuning process, the dense layer parameters $W_0$ remains unchanged and only the low-rank decomposition parameters $\Delta W$ is updated. 
The number of learning parameters is controlled by the rank $r$, we can therefore adjust $r$ during training to achieve a trade-off between computational cost and performance.
According to the comparative experiments in Supplementary Tab.~2, 
we take $r=32$ to achieve excellent performance in most tasks.
To better understand the changes in feature space 
before and after fine-tuning,
we visualized the changes in hidden representation of SAM and SAG as shown in Fig.~\ref{fig:02}c. 
It can be seen that the hidden representation $W_0$ of the model 
before fine-tuning can capture general structural features of seismic images but fail to precisely characterize geobody features, 
whereas the fine-tuned representation $\Delta W$ brings in more detailed geological features.
By merging the $W_0$ and $\Delta W$, the image encoder of
SAG is better adapted to seismic data 
while retaining its original ability to extract common features.
To achieve the optimal performance of fine-tuning, 
we add the LoRA module to each transformer layer of image encoder.
This helps to control the feature space 
at different network depths.

\subsection*{Training and evaluation}\label{sec4sub3}
During the training process, we freeze the pre-trained parameters of the image encoder and only update its corresponding LoRA module. 
To incorporate a new prompt for well logs, the prompt encoder and the mask decoder are fully trained. 
The total trainable parameters amount to only $6.4$ MB, 
which greatly reduces the computational cost compared to retraining.
This model is trained using a weighted sum of cross-entropy loss and Dice loss. 
The AdamW optimizer is employed to optimize the loss function \citep{loshchilov2017decoupled}, 
with an initial learning rate of $1\times10^{-4}$. 
A cosine scheduling strategy is adopted to update the learning rate during training. 
The model is trained for 200 epochs on two V100 (32GB) GPUs, which took about only 10 hours.

In order to simulate actual expert interaction and logging constraints, 
we constructed multiple prompts using masks 
for training, including points, boxes, and well logs. 
To construct point prompts, 
we randomly selected a specific number of points on the binary mask as prompts.
To reduce the model's reliance on too many prompts, 
we reduce the number of points 
as the training process progresses.
Bounding rectangles around the masks were used to generate bounding box prompts. 
To construct prompts from the logs, 
we simulated the lithological segments from well logs as prompts
by extracting multiple straight wells on the binary mask 
and setting the mask outside well logs to zero.
During training, 
we randomly select a single prompt or a combination of various prompts as 
constraints for model prediction.

To evaluate the performance of the SAG model, 
we conducted comparative analysis against the original SAM and CNN-based method (UNet and DeepLabV3+) 
on multi-geobody interpretation tasks involving channels, salt body, paleokarst and strata. 
The UNet is designed with a standard 5-layer UNet model \citep{ronneberger2015u}. 
The DeepLabV3+ adopts ResNet-50 as the backbone network
\citep{chen2018deeplab}. 
For the UNet and DeepLabV3+ models, 
we trained a specific model for each geobody separately.
Due to the fixed network output of CNN methods, 
UNet and DeepLabV3+ cannot achieve strata segmentation in different seismic data.
Therefore, we do not evaluate the UNet and DeepLabV3+ on the task of strata segmentation.
The evaluation results (Fig.~\ref{fig:04} and Supplement tables~1 and ~2) indicated that, 
except for the channel interpretation task, 
the SAG model achieved the best prediction performance for all other tasks.

\subsection*{Loss function}\label{sec4sub4}

We use a weighted sum of the cross-entropy loss function and Dice Loss as the loss function.
Cross-entropy loss is widely applied in semantic segmentation and instance segmentation tasks, 
which is defined by
\begin{equation}
	L_{BCE}=-\frac{1}{N}\sum_{i=1}^{N}[g_i log p_i+(1-g_i)log(1-p_i))],
	\label{equ:lbce}
\end{equation}
where $p_i$, $g_i$ denotes the predicted segmentation and ground truth of pixel i, respectively. 
N is the number of pixels in the image.
But this loss function focuses solely on 
pixel-level classification accuracy and struggles 
to address sample imbalance in the data. 
This sample imbalance issue is prevalent in geobody interpretation tasks. 
Therefore, we incorporated Dice Loss, 
which is defined by
\begin{equation}
	L_{dice}=1-\frac{2\sum_{i=1}^{N}g_i p_i}{\sum_{i=1}^{N}g_i^2+\sum_{i=1}^{N}p_i^2},
	\label{equ:ldice}
\end{equation}
a loss function that has shown excellent performance in medical image segmentation, 
to mitigate the impact of class imbalance in the samples.
The total loss function is defined by
\begin{equation}
	L=\alpha L_{BCE}+(1-\alpha)L_{dice}
	\label{equ:ltt}
\end{equation}
where $\alpha$ is a weight coefficient less than $1$, and its value is empirically taken as $0.3$.
\subsection*{Software}\label{sec4sub6}

All code was developed in Python (3.10) using PyTorch, PyTorch-lightning, 
and Eniops to fine-tune the SAM pre-trained model. 
During the data preparation and processing phase, 
we utilized Python libraries such as OpenCV-Python, Scikit-Image, SciPy, and Matplotlib. 
For the visualization of 3D seismic data and geobodies, 
we employed CIGVIS package \citep{li2024cigvis}. 
Additionally, we implemented a simple interactive 
seismic interpretation software interface based on PyQT5.

%

\bibliography{sn-bibliography.bib}

\bibliographystyle{plainnat}

\clearpage

\section*{Supplementary Information}\label{sec7}

\begin{table}[htb!]
	\begin{center}
		\begin{minipage}{\textwidth}
			\caption{Multi-geobody Interpretation Performance for different methods (Dice Coefficient)}\label{tab1}
			\begin{tabular*}{\textwidth}{@{\extracolsep{\fill}}lcccccc@{\extracolsep{\fill}}}
				\toprule%
				& & \multicolumn{4}{@{}c@{}}{Tasks} \\\cmidrule{3-6}%
				Method & Trainable Parameters & Channel & Salt Body & Paleokarst & Strata  \\
				\midrule
				SAM & 523Mb & 0.4867 & 0.7613 & 0.5725 & 63.38 \\
				UNet & 31.0Mb  & \color[HTML]{0000FE} \textbf{0.8400} & 0,7004 & 0.6712 & -  \\
				DeepLabV3+ & 40.3Mb  & \color[HTML]{FE0000} \textbf{0.8563} & \color[HTML]{0000FE} \textbf{0.8145}  & \color[HTML]{0000FE} \textbf{0.7004}  & - \\
				SAG & 6.4Mb  & 0.7915 & \color[HTML]{FE0000} \textbf{0.8707} & \color[HTML]{FE0000} \textbf{0.7398} & \color[HTML]{FE0000} \textbf{93.13} \\
				\bottomrule
			\end{tabular*}
			\footnotetext{Note: Red and bold values represent the best performance, followed by blue value}
		\end{minipage}
	\end{center}
	\label{tab01:perf}
\end{table}

\begin{table}[htb!]
	\begin{center}
		\begin{minipage}{\textwidth}
			\caption{Multi-geobody Interpretation Performance for SAG with different ranks (Dice Coefficient)}\label{tab1}
			\begin{tabular*}{\textwidth}{@{\extracolsep{\fill}}lcccccc@{\extracolsep{\fill}}}
				\toprule%
				& & \multicolumn{4}{@{}c@{}}{Tasks} \\\cmidrule{3-6}%
				Method & Trainable Parameters & Channel & Salt Body & Paleokarst & Strata  \\
				\midrule
				SAG(rank=8) & 4.7Mb  & 0.7714 & 0.8717  & 0.7242  & 0.9547 \\
				SAG(rank=16) & 5.2Mb  & 0.7754 & \color[HTML]{FE0000} \textbf{0.8785}  & 0.7269  & \color[HTML]{0000FE} \textbf{0.9573} \\
				SAG(rank=32) & 6.4Mb  & \color[HTML]{0000FE} \textbf{0.7915} & \color[HTML]{0000FE} \textbf{0.8707}  & \color[HTML]{0000FE} \textbf{0.7398} & \color[HTML]{FE0000} \textbf{0.9578} \\
				SAG(rank=64) & 8.8Mb  & \color[HTML]{FE0000} \textbf{0.7973} & 0.8687 & \color[HTML]{FE0000} \textbf{0.7516} & 0.9567 \\
				\bottomrule
			\end{tabular*}
			\footnotetext{Note: Red and bold values represent the best performance, followed by blue value.}
		\end{minipage}
	\end{center}
	\label{tab02:rank}
\end{table}

\begin{figure}
	\centering
	\includegraphics[width=\columnwidth]{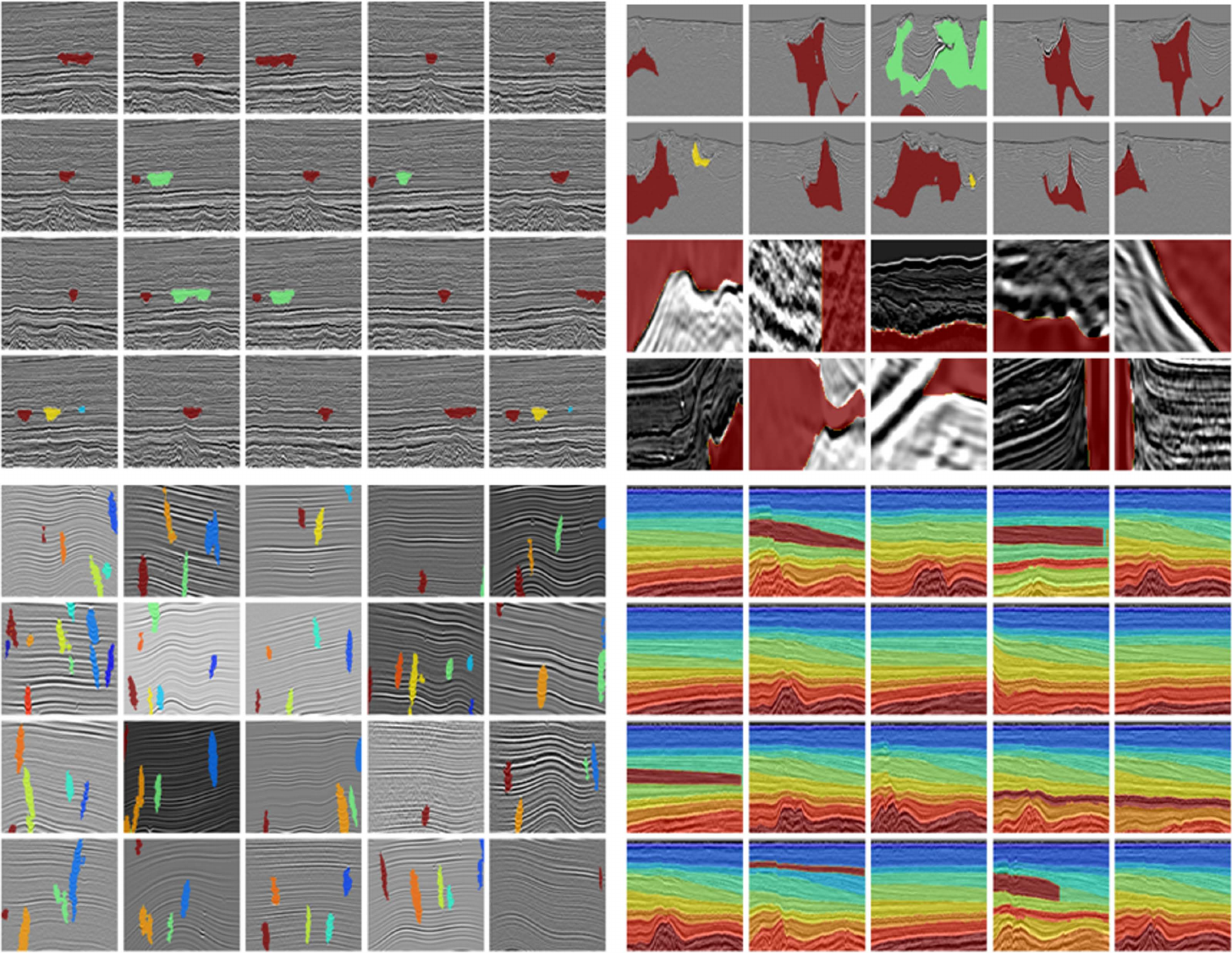}
	\caption{Datasets used for fine-tuning SAG include multiple types of geobodies,
		such as channels, salt bodies, paleokarst
		and strata.
	}
	\label{spfig:01}
\end{figure}

\begin{figure}
	\centering
	\includegraphics[width=\columnwidth]{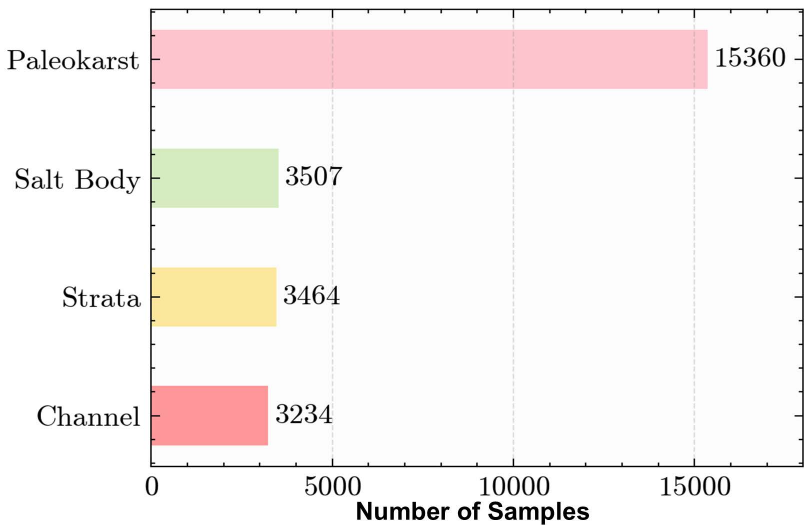}
	\caption{Distribution of seismic image-mask pairs for each type of geobody.}
	\label{spfig:02}
\end{figure}

\begin{figure}
	\centering
	\includegraphics[width=\columnwidth]{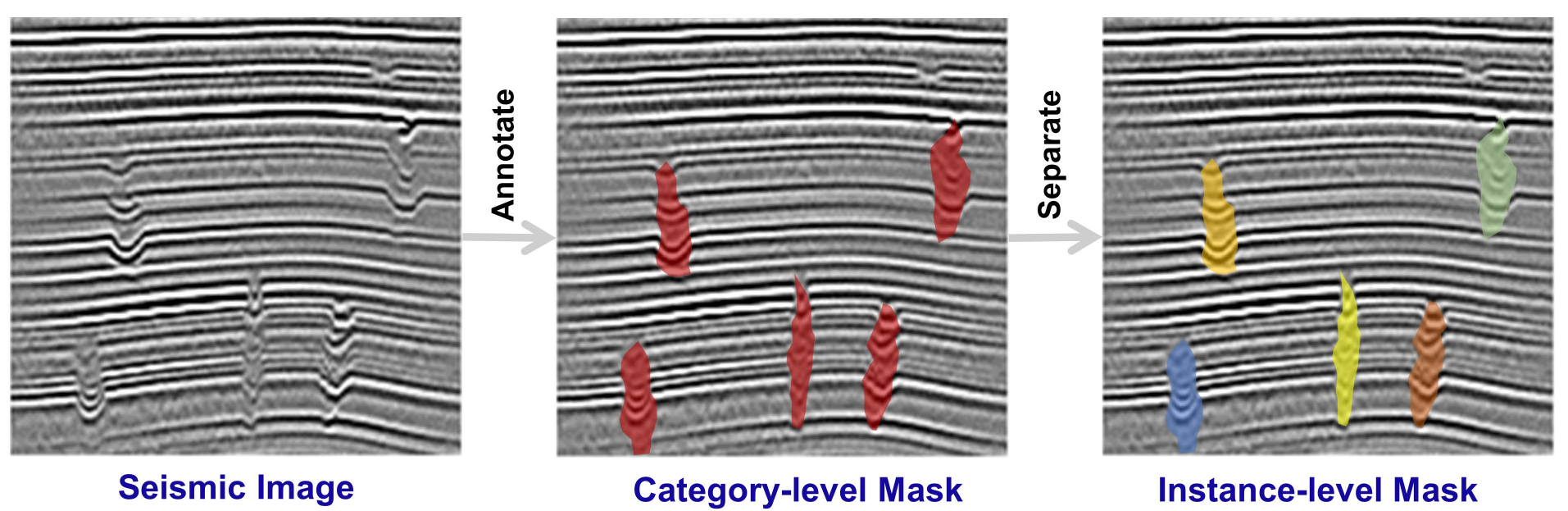}
	\caption{The operation to transform category-level masks into instance-level masks.
		We manually and semi-automatically annotate
		the seismic images to obtain category-level
		masks. 
		Then we use connected component labeling algorithms to convert these masks into
		instance-level masks.
	}
	\label{spfig:03}
\end{figure}

\begin{figure}
	\centering
	\includegraphics[width=\columnwidth]{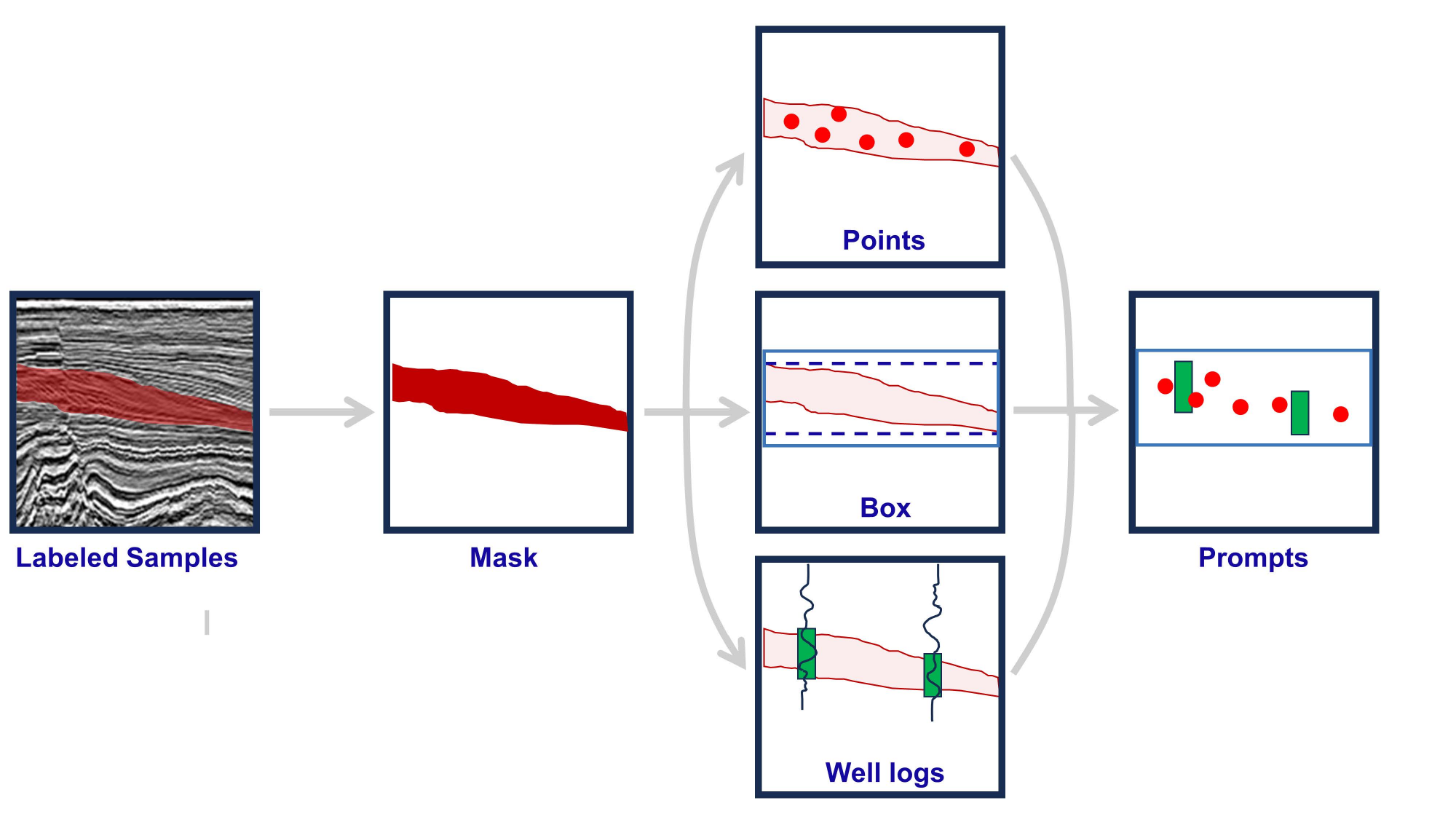}
	\caption{The operation to construct three types of prompts.
		Based on the instance-level masks,
		we can obtain points, bounding boxes and
		well logs as prompts to guide the model
		to segment target geobody.
	}
	\label{spfig:04}
\end{figure}

\begin{figure}
	\centering
	\includegraphics[width=\columnwidth]{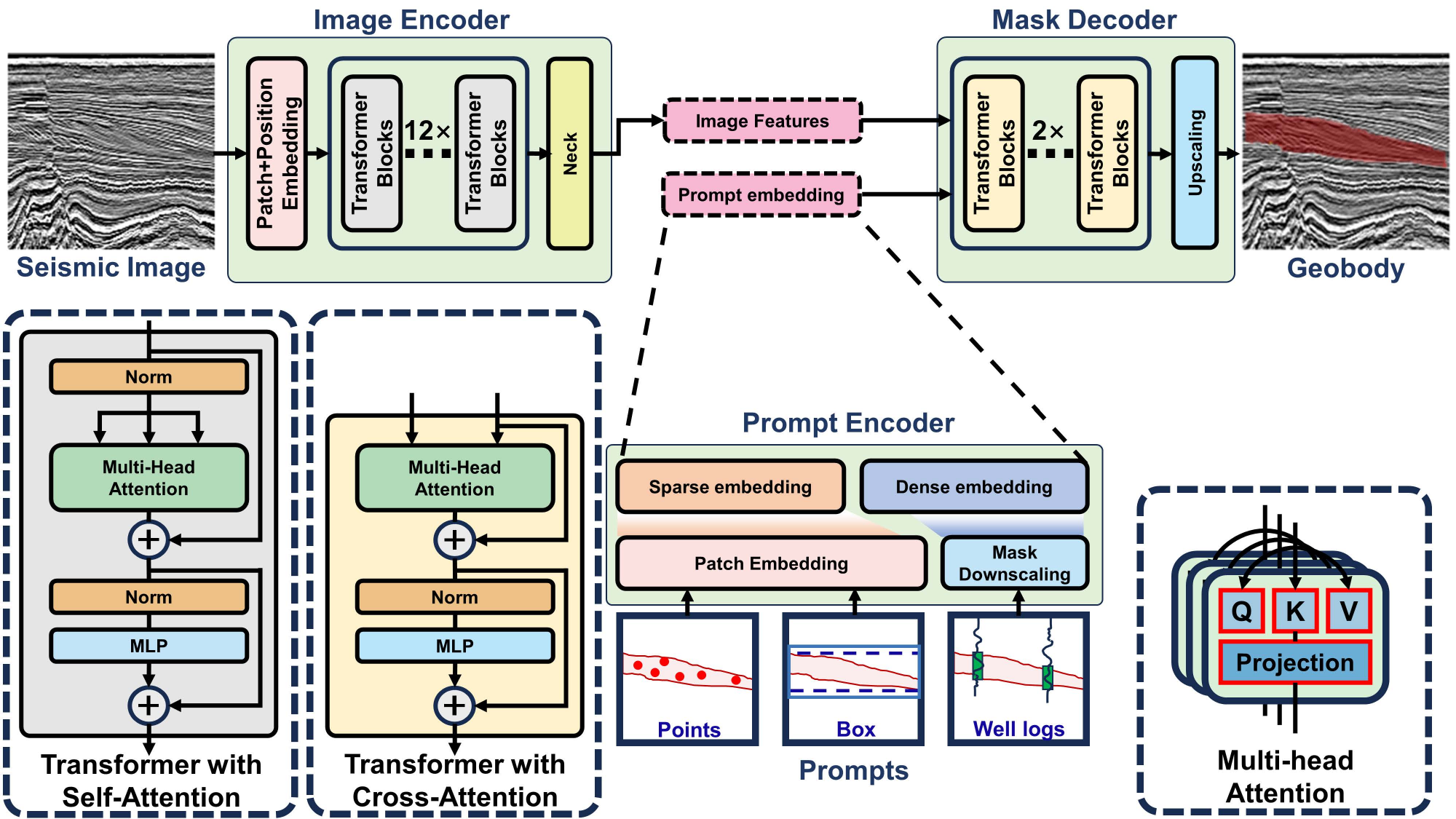}
	\caption{The detailed architecture of SAG.
		SAG contains three modules, image encoder,
		prompt encoder and mask decoder.
		Image encoder module consists of 
		a patch embedding layer, a position embedding
		layer, a self-attention transformer including 12 transformer blocks and a neck layer.
		Prompt encoder module includes 
		a patch embedding layer and a mask downsampling layer.
		Mask decoder module contains 
		a cross-attention transformer with 2 blocks
		and a upsampling layer.
	}
	\label{spfig:05}
\end{figure}

\begin{figure}
	\centering
	\includegraphics[width=0.85\columnwidth]{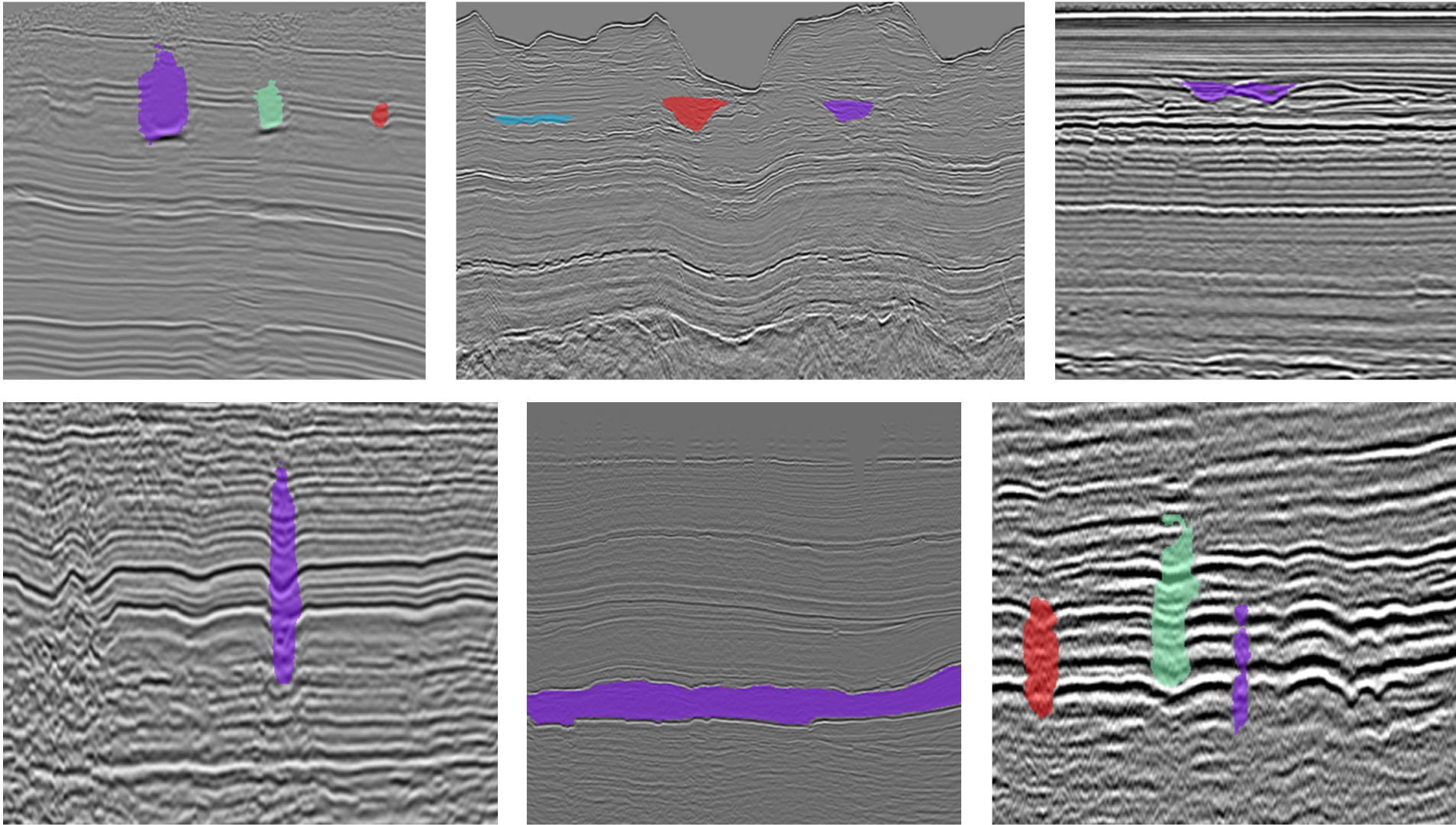}
	\caption{The application of SAG in field seismic surveys.
		We use SAG to interpret multiple geobodies 
		in field seismic data.
		Field seismic examples illustrate SAG
		can achieve universal geobody segmentation
		across different seismic surveys.
	}
	\label{spfig:06}
\end{figure}

\clearpage

\end{document}